\documentclass[%
    aip,
    jcp,
    amsmath,
    amssymb,
    twocolumn,
    superscriptaddress,
    10pt]{revtex4-1}
\usepackage{dcolumn}
\usepackage{graphicx} 
\usepackage{bm}
\usepackage{longtable}
\usepackage[english]{babel}

\begin{document}

\title{Depolarized light scattering and dielectric response of a peptide dissolved in water } 

\author{Daniel R.\ Martin}
\affiliation{Department of Physics, Arizona State University, PO Box 871604, Tempe, AZ 85287-1604 }
   
\author{Daniele Fioretto}
\affiliation{Dipartimento di Fisica, Università di Perugia, via Pascoli, 06123 Perugia, Italy}

\author{Dmitry V.\ Matyushov} 
\affiliation{Department of Physics and Department of Chemistry \& Biochemistry, Arizona State University, PO Box 871604, Tempe, AZ 85287-1604 }
\email{dmitrym@asu.edu} 
 
%\date{\today}

\begin{abstract}
The density and orientational relaxation of bulk water can be separately studied by depolarized light scattering (DLS) and dielectric spectroscopy (DS), respectively. Here we ask the question of what are the leading collective modes responsible for polarization anisotropy relaxation (DLS) and dipole moment relaxation (DS) of solutions involving mostly hydrophobic solute-water interfaces. 
We study, by atomistic molecular dynamics simulations, the dynamics and structure of hydration water interfacing N-Acetyl-leucine-methylamide (NALMA) dipeptide. The DLS response of the solution is consistent with three relaxation processes: bulk water, rotations of single solutes, and collective dipole-induced-dipole polarizability of the solutes, with the time-scale of 130--200 ps. No separate DLS response of the hydration shell has been identified by our simulations.  Density fluctuations of the hydration layer, which largely contribute to the response, do not produce a dynamical process distinct from bulk water. We find that the structural perturbation of the orientational distribution of hydration waters by the dipeptide solute is quite significant and propagates $\sim 3-5$ hydration layers into the bulk. This perturbation is still below that produced by hydrated globular proteins. Despite this structural perturbation, there is little change in the orientational dynamics of the hydration layers, compared to the bulk, as probed by both single-particle orientational dynamics and collective dynamics of the dipole moment of the shells. There is a clear distinction between the perturbation of the interfacial structure by the solute-solvent interaction potential and the perturbation of the interfacial dynamics by the corresponding forces.     
\end{abstract}

\maketitle

\section{Introduction}
\label{sec:1}

The problem of how the static structure and dynamics of water are altered by a solute is at the heart of many problems related to solution chemistry and biology.\cite{Ball:08} The scope of interest here is not limited by the need to understand the overall thermodynamics of hydration. The alteration of the  structure and dynamics of hydration shells surrounding a solute affects both the solute-solute interaction, such as protein-protein and protein-DNA interactions,\cite{Oshima:2011iq} and the kinetics of chemical reactions, such as enzymatic catalysis.\cite{Henzler-Wildman:2007ly} 

Two structural and dynamical characteristics of the interfacial water are at the center of the current focus. The first avenue is to understand the change in the structure of water in the hydration layer compared to the bulk\cite{Lee:84,ChandlerNature:05} and, related to that, the depth of this alteration propagating from the solute's surface into the bulk.\cite{Ebbinghaus:07} The second avenue is the change in the dynamics of water in the hydration layer\cite{Pal:04,Halle:04} and the relative effect of the interface on the single-particle vs.\ collective dynamics.\cite{TielrooijScience:10,Laage:11} The active discussion of these issues is also fueled by the disparity of answers to these questions provided by different experimental techniques on the one hand\cite{Pal:04,Oleinikova:04,Halle:09,Ebbinghaus:07,KhodadadiJPCB:08,Comez:2013kx} and an apparent lack of a universal picture supplied by formal theory and computer simulations on the other.\cite{Giovambattista:2007kx} 

Even the seemingly ``simple'' question of the structure of solute's hydration layers escapes a definitive answer. If any conclusion can be drawn from many years of active research, it is the realization that water can provide many competing structural motives, which can realize with different probability depending on a particular type of the water-solute interaction.\cite{Sharp:10} The interaction with a non-polar solute, which seems to be the simplest problem, has generated an enormous literature devoted to hydrophobicity.\cite{Ball:08,ChandlerNature:05} While many details are still under scrutiny, the general emerging picture is that of two characteristic regimes of hydrophobic hydration, depending on the solute size in the first place.\cite{ChandlerNature:05,Jamadagni:2011tc} Hydrophobic solvation of small solutes does not involve breaking of the water's network of hydrogen bonds, while solutes with their size exceeding $\sim 1$ nm disrupt water's structure, thus resulting in the dominance of surface effects in the hydration thermodynamics.\cite{ChandlerNature:05} Despite this universal trend, the local density of the interfacial water is highly sensitive to the strength of the solute-water attraction, changing from weak de-wetting to a more typical density enhancement when the attraction increases.\cite{Huang:00,Rajamani:05}

The density alteration is not the only effect of the solute on the surrounding water. The orientational structure is also affected. Again, starting with a ``simple'' solute-water configuration, orientational structure spontaneously appears in the interface of water with a non-polar solute or air.\cite{Lee:84,Sokhan:97,Bratko:09} It is driven by the necessity to minimize the free energies of both the dipole and quadrupole moments of water,\cite{Frenkel,StillingerJr:1967tx} which are both significant and carry different symmetries (water's quadrupole is mostly non-axial\cite{Gubbins:84}). This orientational structure, and, in particular, the interfacial quadrupolar density, is responsible for a surface potential of water\cite{Wilson:1988wx,Beck:2013gp,Horvath:2013fe} and the related, but still not entirely understood, asymmetric distribution of positive and negative electrolyte ions in the interfacial region.\cite{Baer:2012uq,Bonthuis:2013fk} 

The structure of water interfacing air or non-polar solutes is still a poor representation of the interface with polar/charged solutes.\cite{Russo:2011hz} In addition to typically denser,\cite{Gerstein:1996ys,Svergun:98,Russo:2011hz} or even collapsed,\cite{Kaya:2013gl} water at hydrophilic surfaces, the orientational structure of water changes dramatically near polar/charged groups compared to non-polar/hydrophobic regions. Different scenarios are possible, but the main qualitative feature is the creation of domains of preferentially oriented waters, consistent with the local electric field produced by the surface groups.\cite{Mondal:2012zr,DMjpcl2:12}  Once the network of hydrogen bonds of bulk water has been disrupted by a sufficiently large solute,\cite{ChandlerNature:05} reorientations of water's multipoles become more feasible and clustering of dipoles, dictated by local fields, takes place.\cite{DMjcp1:11}  Given these dramatic structural changes of interfacial water compared to the bulk, it is almost trivial to say that interfacial dynamics should be different from bulk dynamics. Addressing this question on a more quantitative basis has turned out to be more challenging.\cite{Laage:11,Stirnemann:2011vn}

The main current challenge in understanding the dynamics of hydration shells is to map the properties observed by experiment and produced by computer simulations on particular nuclear modes of hydration waters. Since most observables are weighted averages of different nuclear modes, with different extent of collective behavior,\cite{Castner-Jr.:1998eu} the disparity between interpretations of different experiments is hardly surprising.\cite{Laage:11} Most observations report slowing of the dynamics of first-shell hydration waters compared to the bulk,\cite{Pal:04,Oleinikova:04,Halle:09,Comez:2013kx} but the disagreement still exists on the values of retardation factors and on the spatial extent of slower hydration layers.\cite{Halle:09,Ebbinghaus:07,Comez:2013kx} Among the techniques actively used to probe the dynamics of hydration shells are nuclear magnetic resonance (NMR) (mostly single-molecule rotational dynamics),\cite{Halle:04} Stokes-shift dynamics (collective electrostatic response),\cite{Pal:04,Zhang:07} dielectric spectroscopy (relaxation of the total dipole),\cite{Oleinikova:04,KhodadadiJPCB:08,Perticaroli:2013gc} and depolarized light scattering (DLS)\cite{Perticaroli:10,Comez:2013kx} and time-resolved optical Kerr effect (OKE)\cite{Cho:1993wx,Mazur:2010gr,Mazur:2011ip} (relaxation of optical anisotropy).   

Our focus here is predominantly on the DLS response of solutions. The application of the extended DLS (EDLS) technique to a number of aqueous solutions from carbohydrates,\cite{Lupi:2012bf} to peptides,\cite{Perticaroli:2011gz} to proteins\cite{Perticaroli:10} have produced the picture of increasing retardation of the hydration water with increasing solute size and its hydrophilic character.\cite{Comez:2013kx} Recent molecular dynamics (MD) simulations of EDLS spectra have stressed on the collective nature of the optical anisotropy response,\cite{Lupi:2012bf,Lupi:2012dg} potentially linked to density fluctuations of interfacial waters. The purpose of this study is to further clarify the physical origin of the observed EDLS spectra and to contrast them with the response of the dipole moment of the solute and its hydration shell. 

For bulk water, OKE and DLS techniques are essentially complimentary to dielectric spectroscopy.\cite{Fukasawa:2005cj,Turton:2008kx} Because of small polarizability anisotropy of water, DLS and OKE 
mostly probe dipole-induced-dipole (DID) polarizability\cite{PecoraBook,Stephen:69} related to translational nuclear dynamics.\cite{Madden:1986vh,Mazzacurati:1989vm,Mazzacurati:1990tb,Fecko:2002ys,Turton:2008zr} In contrast, dielectric spectroscopy of bulk water, probing dipole relaxation, is closely linked to rotational diffusion.\cite{Fukasawa:2005cj} These two nuclear modes are separately probed by these two techniques in bulk water. Whether this convenient separation extends to solutions is not entirely clear.\cite{Ernsting:2002ke} Our goal here is to contrast these two types of response by using MD simulations and to clarify the physical mechanisms behind the polarizability anisotropy relaxation detected by the EDLS technique.  

\section{Physical picture}

The common explanation of the optical Kerr effect assumes a molecule with the 2-rank tensor of its electronic (high-frequency) polarizability $\bm{\alpha}$ placed in the external field of electromagnetic radiation. Anisotropy of $\bm{\alpha}$ yields an induced dipole perpendicular the external field, which emits perpendicular polarized scattered light. Rotations of the molecule cause time-dependence of the emitted radiation and, consequently, time dependent OKE signal. OKE spectroscopy thus gives access to time-resolved molecular rotational dynamics, while depolarized light scattering provides the same information in the frequency domain.\cite{PecoraBook}  

The mechanism gives zero OKE or DLS response in the case of isotropically polarizable molecules with values of $\alpha_{\alpha\alpha}$, $\alpha=x,y,z$ in the molecular frame of principal axes diagonalizing $\bm{\alpha}$. However, the DLS signal is nonzero for condensed media of isotropically polarized particles.\cite{PecoraBook,Stephen:69,Greene:1984ly} In that case, the dipole induced at particle $1$ by the external radiation in turn induces the dipole at particle $2$ through the 2-rank tensor of the dipole-dipole interaction $\mathbf{T}_{12}$(DID mechanism). Within this picture, both OKE and DLS spectroscopies report on collective dynamics of highly correlated induced dipoles, and not just on rotational diffusion of individual molecules,\cite{Geiger:1989ul,Elola:2007fk} although rotations and translations are generally coupled in the DID mechanism.\cite{Mazzacurati:1989vm,Murry:1999bh} 

\begin{figure}
  \centering
  \includegraphics*[width=9cm]{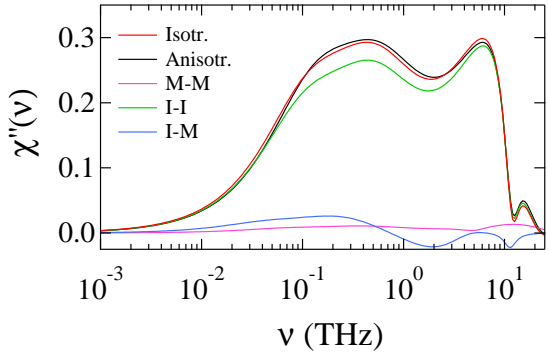}
  \caption{DLS loss spectrum of bulk SPC/E water with anisotropic (black solid line) and isotropic (red solid line) polarizability assigned to the water molecules (see text). The remaining curves show the splitting of susceptibility of anisotropic water into the components arising from molecular polarizability (M-M), from DID induced polarizability (I-I), and from the cross-correlations between the molecular and DID polarizabilities (M-I). The polarizability components of water in the frame of principal axes are:\cite{Scotto:2005} $\alpha_{xx} = 1.408$,  $\alpha_{yy}=1.497$, and $\alpha_{zz}=1.417$ \AA$^3$;  the water molecule is in the $yz$-plane with the $z$-axis set as the axis of rotational symmetry. }
  \label{fig:2}
\end{figure}

It is clear that the DID mechanism should dominate for molecular liquids with weak polarizability anisotropy.\cite{Walrafen:1989bh,Castner:1995fk} This is indeed the case for water characterized by  $\alpha_{xx} = 1.408$,  $\alpha_{yy}=1.497$, and $\alpha_{zz}=1.417$ \AA$^3$ for its principal-axes polarizability when the water molecule is in the $yz$-plane.\cite{Scotto:2005} The results of our calculations of the loss DLS spectrum for pure SPC/E water are shown in Fig.\ \ref{fig:2} and described in more detail below.  

The solid black line in Fig.\ \ref{fig:2} refers to the anisotropic polarizability of water, while the solid red line is obtained assuming isotropic polarizability $\bar\alpha=(1/3)\mathrm{Tr}[\bm{\alpha}]$ assigned to all principal polarizability components. As is seen, the DLS signals are nearly identical for isotropically and anisotropically polarizable water. The induced DID component of the response, marked ``I-I'' in Fig.\ \ref{fig:2}, dominates the DLS spectrum. The relevant question here is the dynamics of which nuclear modes are displayed by the DLS response. Given that dipole-dipole interactions are sensitive to intermolecular distance, a reasonable conjecture is that either single-particle translational dynamics or the dynamics of collective density fluctuations are measured in the first place.  The DLS response of isotropically polarizable liquids can in fact be described in terms of the dynamic density structure factors (collective density fluctuations),\cite{Stephen:69} while a small fraction of the OKE response can still be assigned to rotational diffusion.\cite{Winkler:2000vy} The dominance of translational modes and of the DID mechanism in the DLS response are preserved for the water component of the mixture as well.\cite{Elola:2007fk,Lupi:2012bf}           

The situation becomes significantly more complex for solutions. Solutes often carry anisotropic polarizability, and their DLS response can naturally be assigned to their rotational diffusion.\cite{Elola:2007fk,Fioretto:2007jl} However, in addition to the typically slowest DLS band assigned to solute's rotations, bands intermediate in frequency between the solute and water have been detected in the solution spectra and assigned to hydration water shells,\cite{Lupi:2012bf,Comez:2013kx} with the intensity of the intermediate DLS peak associated with the number of hydrated hydroxide groups.\cite{Lupi:2012dg}  

The solute we study here, N-Acetyl-leucine-methylamide (NALMA), is considered to be hydrophobic, with no specific interactions with waters. It presents therefore an opportunity of a simpler system, compared to sugars, for which relative contributions from different relaxation processes to the DLS signals can be disentangled by MD simulations. Aqueous solution of NALMA have been studied by a number of experimental techniques including neutron scattering,\cite{Russo:2004uq,Russo:2011hz,Russo:2013iv} terahertz absorption,\cite{Born:2009zr} NMR,\cite{Qvist:2009kx} OKE,\cite{Mazur:2010gr} and EDLS\cite{Petricaroli:11} spectroscopies. Classical MD simulations of aqueous NALMA solutions have also been reported.\cite{Murarka:2007ly,Qvist:2009kx,Russo:2011hz,Russo:2013iv} 

A fairly simple decomposition of the DLS signal indeed emerges from our simulations. We do not detect the 50-60 GHz component observed for hydrated sugars,\cite{Fioretto:2007jl,Gallina:2010de,Lupi:2012bf} but find two slower modes which can be assigned to single-molecule NALMA rotations and collective DID relaxation of NALMAs in solution. No slowing down of the hydration layers is seen either in the DLS response of solutions or in the dipolar relaxation of hydration shells. The hydration shell nevertheless displays a significant orientational structure propagating $\simeq 3-5$ hydration layers into the bulk. 

\section{Formalism} 
The dipolar polarizability of the solution $\bm{\Pi}$ is given as a sum of the
molecular polarizability $\bm{\Pi}^{\text{M}}$ and the polarizability $\bm{\Pi}^{\text{I}}$ caused by interactions between induced molecular dipoles\cite{Elola:2007fk}
\begin{equation}
\label{eq:1}
\bm{\Pi} = \bm{\Pi}^{\text{M}} + \bm{\Pi}^{\text{I}} , 
\end{equation}
where 
\begin{equation}
\bm{\Pi}^{\text{M}}=\sum_i^{N_0} \bm{\alpha}_i^{(0)} + \sum_i^{N_s} \bm{\alpha}_i
\end{equation}
adds $N_0$ polarizabilities of the solutes and $N_s$ polarizabilities of the solvent molecules. The induction component in Eq.\ \eqref{eq:1} is not pair-wise decomposable since it arises from all chains of dipole-dipole interactions between the induced electronic dipoles.\cite{Wertheim:1979ty,SPH:81,Cao:1993uq} Full classical description of this term requires self-consistent calculations of all induced dipoles in the sample. A perturbation simplification of this expensive algorithm, which was shown to be accurate in application to typical dense polarizable liquids,\cite{Geiger:1989ul,Elola:2007fk} involves only the sum over pairwise dipole-dipole coupling terms. One gets, for instance, for the solvent component
\begin{equation}
\label{eq:2}
\bm{\Pi}_s^{\textbf{I}} = \sum_{i\ne j}^{N_s} \bm{\alpha}_i\cdot \mathbf{T}_{ij}\cdot \bm{\alpha}_j ,
\end{equation}
where $\mathbf{T}$ is the dipolar tensor. The polarizability of water is the sum of molecular and induced polarizabilities, $\bm{\Pi}_s= \bm{\Pi}_s^{\text{M}} + \bm{\Pi}_s^{\text{I}}$. 

Dipolar polarizabilities $\bm{\alpha}_i$ are assigned to oxygen atoms in the case of water. This simplified scheme cannot be used for extended solutes, and a formalism accounting for a distributed polarizability is required. We have used here, following Ladanyi and co-workers,\cite{Elola:2007fk,Lupi:2012dg} the Thole formalism that assigns an induced point dipole to each atom in a multi-atom molecule.\cite{Thole:1981,Duijnen:1998} Since mutual dipolar interactions between induced atomic dipoles need to be accounted for, the effective atomic polarizabilities of the solute atoms are obtained by inverting the matrix containing self polarizabilties and their dipole-dipole couplings at each configuration of the solute molecule along the simulation trajectory. The procedure is explained in more detail in the Supplementary Material (SM),\cite{supplJCP} but its outcome is a $3n_a\times 3n_a$ matrix obtained by inverting the combination of $3\times 3$ site atomic polarizabilities with dipole-dipole interactions between them. This matrix then defines the atomic dipolar polarizability $\bm{\alpha}_{ip}^{(0)}$ at each site $p$ of solute $i$, where $p=1,\dots,n_a$ runs over $n_a$ atoms of the solute ($n_a=31$ for NALMA).   

Since the dipole-dipole coupling between the induced dipoles at the atomic sites of the solute are taken into account to all orders by the matrix inversion, there is no need in accounting for induced polarizability of the solute sites. Induced polarizabilities, however, arise from dipole-dipole interactions between the oxygens of water and the atomic sites of the solutes. The solute-solvent induced polarizability therefore reads
\begin{equation}
\label{eq:3}
\bm{\Pi}_{0s}^{\text{I}} = \sum_{i=1}^{N_0}\sum_{p=1}^{n_a} \sum_{j} \bm{\alpha}_{ip}^{(0)}\cdot \mathbf{T}_{ip,j}\cdot \bm{\alpha}_j, 
\end{equation}
where $\mathbf{T}_{ip,j}$ is the dipolar tensor connecting site $p$ of solute $i$ (out of $N_0$ solutes overall) to the oxygen atom of water $j$. Similarly, the induced polarizability from the solute-solute dipolar interactions reads as
\begin{equation}
\label{eq:3-1}
\bm{\Pi}_{00}^{\text{I}} = 2\sum_{i=1}^{N_0}\sum_{j<i}\sum_{p,q=1}^{n_a}  \bm{\alpha}_{ip}^{(0)}\cdot \mathbf{T}_{ip,jq}\cdot \bm{\alpha}_{jq}^{(0)},
\end{equation}
where $\mathbf{T}_{ip,jq}$ is the dipolar tensor connecting site $p$ of 
solute $i$ with site $q$ of solute $j\ne i$. Reaction-field corrections for the finite size of the simulation box\cite{Allen:96} were included in all dipolar tensors used to calculate the dipolar interactions from the simulation trajectories (see SM\cite{supplJCP}).  

The overall polarizability $\bm{\Pi}_0$ of the solutes is the sum of $N_0$ molecular polarizabilities calculated from the Thole inversion,\cite{Thole:1981,Duijnen:1998} with their mutual induced polarizabilities
\begin{equation}
\bm{\Pi}_0 =  \bm{\Pi}_{0}^{\text{M}} + \bm{\Pi}_{00}^{\text{I}} ,
\label{eq:3-2}
\end{equation}
where
\begin{equation}
\bm{\Pi}_{0}^{\text{M}} = \sum_{i=1}^{N_0} \bm{\alpha}_i^{(0)} .
\label{eq:3-3}
\end{equation}
The total polarizability of the sample becomes
\begin{equation}
\label{eq:4}
\bm{\Pi} = \bm{\Pi}_0 + \bm{\Pi}_s + \bm{\Pi}_{0s}^{\text{I}} . 
\end{equation}

The trajectory of the solution polarizability is used to calculate the time autocorrelation function based on its off-diagonal components 
\begin{equation}
\label{eq:5}
 C^{\Pi}(t) \propto \sum_{\alpha \ne \beta} \langle \delta\Pi_{\alpha\beta}(t) \delta\Pi_{\alpha\beta}(0) \rangle, 
\end{equation}
where $\alpha=x,y,z$ are the Cartesian components of the 2-rank Cartesian tensor. The time correlation function is used to define the OKE response function $\chi(t)= -\beta \dot C^{\Pi}(t)$, where $\beta=1/(k_{\text{B}}T)$ is the inverse temperature. The response function can be Fourier-Laplace transformed to connect to the frequency Fourier transform $C^{\Pi}(\omega)$ of $C^{\Pi}(t)$ via the classical limit of the fluctuation-dissipation relation\cite{Hansen:03} (see SM\cite{supplJCP} for more detail)
\begin{equation}
\label{eq:6}
\chi''(\omega) = (\beta\omega/2) C^{\Pi}(\omega) . 
\end{equation}

\section{Simulation Protocol} 
A single NALMA molecule was geometrically optimized and partial
atomic charges were assigned using density functional calculation with the 6-311G basis set (Gaussian'09\cite{g09}). The torsion and Lennard-Jones potentials for atomic sites were taken from CHARMM'22.\cite{charmm22} Quantum calculated geometries were used as the starting point in a 10000 step energy minimization, followed by 10--20 ns NPT equilibration in the cubic simulation box filled with NALMA molecules and SPC/E waters. NPT simulations were performed using the Langevin pressure and temperature controls with a piston pressure of 1.0 atm, piston decay of 50 fs, damping coefficient of 5 ps$^{-1}$, and piston period of 100 fs. The MD simulations were done using the NAMD 2.8 simulation package.\cite{namd} The number of waters and NALMAs at different concentrations studied here are listed in Table \ref{tab:1}.

\begin{table}
\begin{ruledtabular}
\centering
  \caption{Number of waters $N_s$ and number of NALMA solutes $N_0$ in the simulation cell. Also listed are the molar concentrtion of the solute $c_0$, the side length $L$ of the cubic simulation box, and the simulation time $t_{\text{sim}}$. }
  \label{tab:1}
  \begin{tabular}{lccccc}
     Label & $N_s$ & $N_0$ & $c_0$, M & $L$,\AA & $t_{\text{sim}}$, ns\\
    \hline
     C0 & 11721 &  1  &      &  72    & 400\\
     C1 & 36269 &  80 & 0.12 & 104 & 185 \\
     C2 & 39430 & 225 & 0.30 & 108 & 100 \\
     C3 & 37122 & 330 & 0.45 & 107 & 100 \\
  \end{tabular}
\end{ruledtabular}  
\end{table}

Production runs were performed for bulk SPC/E water (20 ns, $~58$ \AA\ simulation box containing 6841 waters) and NALMA solutions (Table \ref{tab:1}) in the NVE ensemble. All simulations were done at 300 K. Temperature was maintained by performing 5 ns NPT runs after each 20 ns NVE run. Periodic boundary conditions were used for all simulations. The long-ranged electrostatics was handled with the electrostatic cutoff of 12 \AA\ and the particle mesh Ewald algorithm\cite{spme:95} provided by NAMD. For NALMA solutions, the saving frequency of 0.05 ps was applied to the first 20 ns of simulations, followed by the saving frequency of 0.5 ps for the rest of the trajectory. 

\begin{figure}
  \centering
  \includegraphics*[width=9cm]{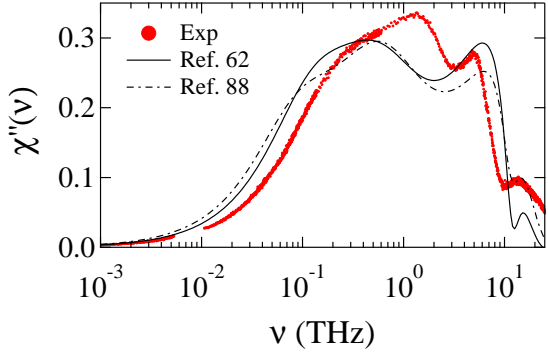}
  \caption{DLS loss spectrum of bulk SPC/E water (black solid line). The red points are data from the EDLS experiment.\cite{Petricaroli:11} Also shown (dash-dotted line) is the DLS loss spectrum produced with the polarizability matrix used in Refs.\ \onlinecite{Sonoda:2005bm,Lupi:2012bf}: $\alpha_{xx} = 1.04$,  $\alpha_{yy}=1.00$, and $\alpha_{zz}=1.17$ \AA$^3$. }
  \label{fig:3}
\end{figure}

\section{Polarizability response}

\subsection{Bulk water} 
The DLS response of bulk water was calculated from separate MD simulations of SPC/E water at 300 K. The data were analyzed by assigning anisotropic polarizability\cite{Scotto:2005} to water, with the principal-axes components listed above and in the caption to Fig.\ \ref{fig:2}. The normalized  time correlation function $\phi_s(t)=C^{\Pi}(t)/C^{\Pi}(0)$ was fitted to a linear combination of a Gaussian decay, a damped harmonic oscillator, and two exponential Debye terms. Two restrictions, $\phi(0)=0$ and $\dot \phi(0)=0$, were additionally imposed, resulting in the following fitting function (see SM\cite{supplJCP}) 
\begin{equation}
\label{eq:7}
\phi(t) = e^{-\omega_g^2 t^2/2} + \sum_{i=1}^{2}  B_i g_i(t),
\end{equation}
where 
\begin{equation}
\label{eq:7-1}
g_i(t) = e^{-\alpha_i t} +\left( \alpha_i \tau_h - 1\right) e^{-\omega_g^2 t^2/2} - \alpha_i \tau_h e^{-t/\tau_h}\cos\omega_h t .
\end{equation}
Eight free parameters obtained by fitting Eqs.\ \eqref{eq:7} and \eqref{eq:7-1} to MD data are listed in Table S1 in SM.\cite{supplJCP}  Larger sets of fitting parameters, as well as two damped harmonic oscillators, have been tried, but could not provide decisively better representation of the data.

\begin{figure}
  \centering
   \includegraphics*[width=9cm]{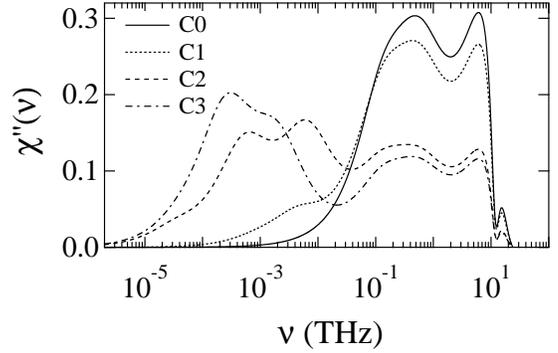}
   \caption{DLS susceptibilities from MD simulations for solution concentrations
    listed in Table \ref{tab:1}. }
  \label{fig:4}
\end{figure}

The DLS loss spectrum of anisotropically polarizable SPC/E water is compared in Fig.\ \ref{fig:3} to the experimental spectrum of bulk water.\cite{Petricaroli:11}  We also show in the figure the results of the same analysis applying somewhat reduced values of the water polarizability used in previous modeling of OKE and DLS spectra.\cite{Sonoda:2005bm,Lupi:2012bf} The agreement between the simulations and experiment is only semi-quantitative, which can be traced back to deficiencies of the force field and to the perturbation approximation of the DID polarizability used in the analysis (Eq.\ \eqref{eq:2}). Given these deficiencies, the main goal of our analysis is not to quantitatively match the measurements, but to gain qualitative insights into different physical processes and nuclear modes contributing to the DLS response. 

\subsection{NALMA solution}
Simulations at 300 K were carried out for a single NALMA solute in the simulation cell (labeled as C0) and three aqueous NALMA solutions with finite numbers of NALMAs in the simulation cell (Table \ref{tab:1}). In contrast to previous MD simulations with the OPLS force field used for NALMA and covering much shorter simulation trajectories of more concentrated solutions,\cite{Murarka:2007ly,Murarka:2007fk} our solutions with concentrations $c_0$ marked as C2 and C3 in Table \ref{tab:1} are unstable to solute aggregation. This is seen from the growing slow relaxation tail in the susceptibility functions in Fig.\ \ref{fig:4}, as well as from the corresponding solute-solute pair distribution functions presented in SM.\cite{supplJCP} 

These observations are roughly consistent with experiment, which shows NALMA aggregation at $c_0>1$ M in pH 7 solution;\cite{Russo:2004uq} reaching higher concentration requires acidic pH.\cite{Born:2009zr,Mazur:2010gr} The disparity between different experimental reports might therefore be partially related to differences in the degree of peptide's protonation. Further, the OKE relaxation time of NALMA solutions is a linear function of the peptide concentration up to $c_0\simeq 0.5$ M and increases non-linearly for more concentrated solutions.\cite{Mazur:2010gr}  Similarly, the NMR relaxation rate was found to change linearly with NALMA's concentration up to 0.22 M at pH 4.5.\cite{Qvist:2009kx} 

\begin{figure}
  \centering
  \includegraphics*[width=9cm]{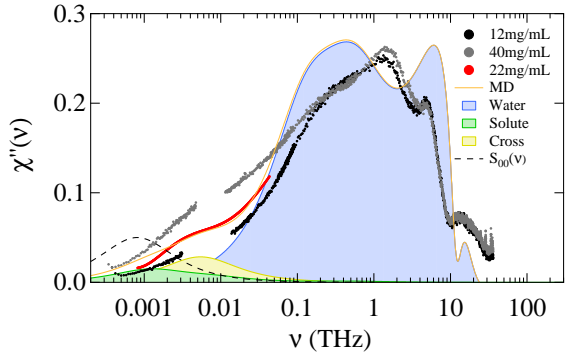}
  \caption{DLS loss spectrum (solid orange line, C1 solution, Table \ref{tab:1}) and its separation into the bulk water, solutes in the solutions, and cross correlations (Eq.\ \eqref{eq:8}, Table \ref{tab:2}). The concentration of NALMA in simulations is 22 mg/ml. Points show experimental DLS spectra obtained at 12 mg/ml (black) and 40 mg/ml (grey),\cite{Petricaroli:11} the red line is the interpolation between the two sets of data to the MD concentration of 22 mg/ml. Both simulation and experimental spectra are equally normalized.  The dashed line indicates the loss spectrum of the partial solute-solute dynamic structure factor $S_{00}(k^*,\omega)$ [Eq.\ \eqref{eq:10-2}] calculated at the wave vector $k^*=2\pi/\langle r \rangle $ corresponding to the average distance $\langle r\rangle= 24.14$ \AA\ between the NALMAs in solution. The height of this loss spectrum is arbitrary rescaled to match the scale of the plot.  }
  \label{fig:5}
\end{figure}

Our main conclusions regarding the DLS response are based on C0 and C1 solutions with no aggregation  (Table \ref{tab:1}). We, however, also present the results for C2 and C3 solutions to indicate possible signatures of aggregation and, in particular, slow dynamics of water trapped in the aggregates. Further, C1 concentration ($\sim 22$ mg/ml) falls in the concentration range, 12--100 mg/ml, studied experimentally.\cite{Petricaroli:11} An interpolation between the closest sets of experimental data to C1 concentration is shown in Fig.\ \ref{fig:5}. Note a good match between experimental and theoretical DLS intensities in the low-frequency part of the spectrum where deficiencies of the force fields are expected to be less pronounced.

\begin{table}
\begin{ruledtabular}
\centering
  \caption{Muti-exponential fit of time correlation functions from MD simulations; relaxation times are in ps. The label in parentheses indicates the solute concentration as listed in Table \ref{tab:1}.   }
  \label{tab:2}
  \begin{tabular}{lcccccc}
     Function & $A_1$ & $A_2$ & $A_3$ & $\tau_1$ & $\tau_2$ & $\tau_3$\\
    \hline
     $C_0^{\Pi}(t)$ (C0) & 0.17 & 0.83 &  & 0.4 & 26 & \\
     $C_0^{m}(t)$ (C0) & 0.20 & 0.80 & & 7.3 & 73 & \\
\hline     
  $C_0^{\Pi}(t)$ (C1) & 0.14 & 0.86 &  & 0.5 & 27 &  \\
  $C^{m}_0(t)$ (C1) & 0.27 & 0.73 & & 11.3 & 84 & \\
  $\tilde C_0^{\Pi}(t)$ (C1) & 0.06 & 0.24 & 0.70 & 0.3 & 20 & 136 \\
     $C^{\Pi}(t)$ (C1)\footnotemark[1] & $-3.0\times10^{-4}$ & 0.06 & 0.90  & 0.03 & 29 & \\
     $F_{00}(k^*,t)$ (C1)\footnotemark[2] &      &      & 1.0 &  &  & 196 \\
  \end{tabular}
\end{ruledtabular}
\footnotetext[1]{$A_3=B_1$ is the water component in the correlation function [Eq.\ \eqref{eq:8}].  }
\footnotetext[2]{Partial intermediate scattering function of solutes in solution [Eq.\ \eqref{eq:10-2}] calculated at $k^*=2\pi/\langle r\rangle$ corresponding to the average distance $\langle r\rangle = 24.14$ \AA\ between the solutes. }
\end{table}

\subsection{DLS response}
We start our discussion of the DLS response of solutions with the case of a single solute in the simulation cell (C0 in Table \ref{tab:1}). Two time autocorrelation functions are used to study the solute dynamics. The first autocorrelation function 
\begin{equation} 
\label{eq:9}
C_0^{\Pi}(t) \propto  \sum_{\alpha \ne \beta} \langle \delta\Pi_{0,\alpha\beta}^{\text{M}}(t) \delta\Pi_{0,\alpha\beta}^{\text{M}}(0) \rangle 
\end{equation}
describes the relaxation of the single solute polarizability $\bm{\Pi}_0^{\text{M}}=\bm{\alpha}^{(0)}$, also including the  mutually induced atomic dipoles (DID mechanism). The second autocorrelation function is that of the solute dipole $\mathbf{m}_0(t)$, which mostly reflects its rotational diffusion,\cite{Debye:29} 
\begin{equation}
\label{eq:10}
C^{m}_0(t) \propto \langle \delta\mathbf{m}_0(t)\cdot \delta\mathbf{m}_0(0)\rangle .  
\end{equation}
Both functions were fitted to a two-exponential decay and the results are listed in Table \ref{tab:2}.    
   
The solute dipole correlation function $C_0^{m}(t)$ is the contraction of the first-rank tensor, while $C_0^{\Pi}(t)$ represents the dynamics of the second-rank tensor. If the same rotational diffusion contributes to both of them, the corresponding relaxation times should follow the simple scaling $\tau_{\Pi}=\tau_m/3$.\cite{PecoraBook,Perticaroli:2013gc} This is indeed the case, within the uncertainties of fitting, for the slowest relaxation times obtained from fitting  $C_0^{m}(t)$ and $C_0^{\Pi}(t)$ of both C0 and C1 solutions (see $\tau_2$ for C0 and C1 in Table \ref{tab:2}). The slowest relaxation time obtained here is also consistent, within concentration differences between simulations and measurements, with the slowest relaxation peak measured by the EDLS.\cite{Perticaroli:2011gz} The experimental EDLS peak was also assigned to rotational diffusion based on its dependence on temperature and solution viscosity.  

\begin{figure}
\includegraphics*[width=7cm]{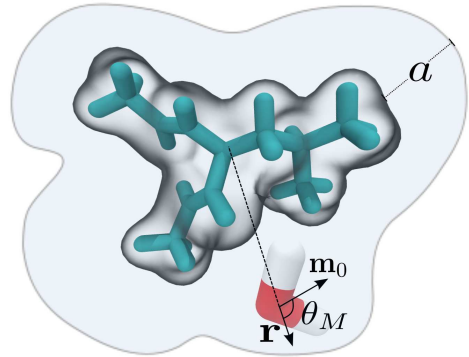}
\caption{Cartoon of NALMA in solution. Shown is the hydration layer of thickness $a$ and the angle $\theta_M$ between the water's dipole moment and the radial vector drawn from NALMA's center of mass.} 
\label{fig:6}
\end{figure}

For the C1 configuration we report the correlation function of the entire polarizability of all 80 solutes in the simulation box
\begin{equation}
\tilde C_0^{\Pi}(t) \propto N_0^{-1}\sum_{\alpha\ne\beta} 
   \left\langle \delta\Pi_{0,\alpha\beta}(t) \delta\Pi_{0,\alpha\beta}(0) \right\rangle ,
\label{eq:10-1}
\end{equation}
where $\bm{\Pi}_0$ is defined by Eq.\ \eqref{eq:3-2}. In contrast to $C_0^{\Pi}(t)$ (Eq.\ \eqref{eq:9}) and $C_0^{m}(t)$ (Eq.\ \eqref{eq:10}), this correlation function cannot be fitted by two exponents since an additional slow relaxation appears in the decay (Table \ref{tab:2}). The dynamics of the solute-solute density fluctuations help to identify the physical origin of this relaxation component. The partial intermediate scattering function 
\begin{equation}
\label{eq:10-2}
F_{00}(k,t) = N_0^{-1} \sum_{i,j=1}^{N_0} e^{i(\mathbf{r}_{0i}(t)-\mathbf{r}_{0j}(0))\cdot \mathbf{k}}
\end{equation}
describes the dynamics of density fluctuations of solutes with coordinates $\mathbf{r}_j(t)$ probed at the wave vector $\mathbf{k}$. The relaxation at $k^*=2\pi/\langle r\rangle$, corresponding to the average distance between the solutes $\langle r\rangle$, is single-exponential (Table \ref{tab:2}). The relaxation time of $F_{00}(k^*,t)$ matches well the slowest relaxation time of the DLS response (Fig.\ \ref{fig:5}). Since $F_{00}(k,t)$ describes density fluctuations of the solutes in solution, the slowest relaxation component in $\tilde C_0^{\Pi}(t)$ is assigned to the DID polarization relaxation caused by solute translations. 

The dynamics of polarization anisotropy relaxation of NALMAs in solution are highly cooperative. This is seen from  a significant compensation between the self correlation functions of the solutes with their cross-correlations. The correlation function $\tilde C^{\Pi}_0(t)$ in Eq.\ \eqref{eq:10-1} can be split into the self-correlation contribution $C^{\Pi}_0(t)$ and all possible cross-correlations lumped into $C^{\text{cross}}_0(t)$: $\tilde C^{\Pi}_0(t) = C^{\Pi}_0(t)+ C^{\text{cross}}_0(t)$. At $t=0$ one gets from simulations: $C^{\Pi}_0(0)=142$ \AA$^6$, $\tilde C^{\Pi}_0(0)=2.3$ \AA$^6$, and $\tilde C^{\text{cross}}_0(0)=-139.7$ \AA$^6$. There is a nearly 98\% cancellation between self and cross components. This type of cancellation is often observed for the dipolar response of polar liquids.\cite{DMjcp2:04} However, its appearance in this problem testifies to a high extent of correlation between NALMAs' induced dipoles in solution. Judging from the polarization anisotropy relaxation, the solution is highly non-ideal in this concentration range. 

Slow relaxation components seen for the solute alone are also present in the time correlation function $C^{\Pi}(t)$ of the entire polarizability of the solution [Eq.\ \eqref{eq:5}]. They are therefore attributed to the solute component of the solution. The next question is whether there are new relaxation processes in the solution, which cannot be assigned to separate bulk and solute dynamics. In order to address this question, we represent the time correlation function of the entire solution polarizability as a weighted sum of the bulk relaxation $\phi_s(t)$, relaxation of all solutes in the solution $\phi_0(t)=\tilde C_0^{\Pi}(t)/\tilde C_0^{\Pi}(0)$, and the rest, lumped into $\phi_{\text{cross}}(t)$
\begin{equation}
\label{eq:8}
  \phi(t)  =  B_1 \phi_{s}(t) + B_2 \phi_0(t) + \phi_{\text{cross}}(t) .
\end{equation}
In addition, constraints $\phi(0)=1$ and $\dot \phi(0)=0$ are imposed on $\phi(t)$. 
The bulk function $\phi_{s}(t)$ is given by Eq.\ \eqref{eq:7} (Fig.\ \ref{fig:3}, also see Fig.\ S1 and Table S1 in SM\cite{supplJCP}). The subscript ``cross'' in the last term in Eq.\ \eqref{eq:8} reflects the expectation that at least cross-correlations between bulk water and solutes will contribute to this term. An independent dynamic process, such as new dynamics of hydration shells, is expected to contribute to  $\phi_{\text{cross}}(t)$ as well. It was modeled by two decaying exponents ($N_e=2$)
\begin{equation}
\phi_{\text{cross}}(t) = \sum_{i=1}^{N_e} A_i e^{-t/\tau_i} .
\label{eq:8-1}
\end{equation}

The fitting of MD results to Eqs.\ \eqref{eq:8} and \eqref{eq:8-1} shows  only about 6\% of the intensity from $\phi_{\text{cross}}(t)$ (yellow in Fig.\ \ref{fig:5}). The response is therefore mostly a linear addition of the bulk and solute components. The solute component (green in Fig. \ref{fig:5}) is the composite peak (see $\tilde C_0^{\Pi}(t)$ entry in Table \ref{tab:2}) including both the solute rotations (faster) and their density fluctuations (slower). Further, the exponential relaxation times of the cross term in Eq.\ \eqref{eq:8-1} are listed in Table \ref{tab:2}. They match very well the solute rotational diffusion and bulk water dynamics. This match strongly suggests that the origin of this component is in cross correlations between bulk water and the solute rotations. No new relaxation process could be identified from this analysis.

\begin{figure}
\includegraphics*[width=8cm]{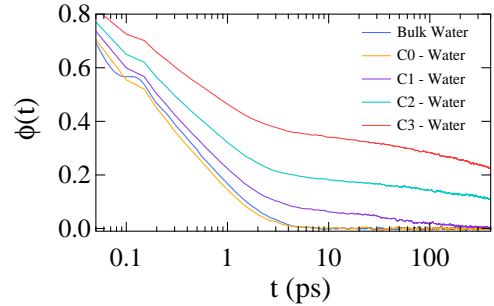}
\caption{Normalized time DLS correlation function of the polarizability of water $\bm{\Pi}_s(t)$ in solutions indicated in the plot. }
\label{fig:7}
\end{figure}

In order to further support these conclusions, we have looked at the dynamics of the water component of the solution. The corresponding relaxation functions (Fig.\ \ref{fig:7}) clearly show growing slow relaxation tails when the concentration of NALMAs is increased. Figure \ref{fig:7} also presents the results for the aggregated solutions C2 and C3 (Table \ref{tab:1}), which develop much slower, 2--3 ns, relaxation components. These relaxation times reflect water trapped in the ``nanopools'' within NALMA aggregates.\cite{Giammanco:2012ys,Zhang:2013ly} 

To understand the origin of slow tails in non-aggregated C0 and C1 solutions, we have analyzed the corresponding time correlation functions with Eqs.\ \eqref{eq:8} and \eqref{eq:8-1}. The question here is whether slow components represent waters moved by the solutes or some additional relaxation processes, such as those originating from the hydration layer. Fitting $\phi(t)$ of C1 solution yields $A_1 = -2.4\times10^{-4}$ and $A_2 = 0.026$, with the corresponding relaxation times $\tau_1=0.02$ and $\tau_2=44$ ps in Eq.\ \eqref{eq:8-1}. In other words, 97\% of the water correlation function is a direct weighted addition of the water and solute correlation functions. The outcome is very similar to the results of fitting the correlation function of the entire polarizability of the solution. It suggests that slow tails appearing in the correlation functions of the water sub-ensemble can be assigned to waters moved by the dissolved solutes.

The slowest dynamics ($\sim 136$ ps) seen for the solutions are not observed for the C0 configuration with a single solute in the simulation cell. This is consistent with our suggestion that only two dynamic processes contribute to the infinite dilution signal: solute rotations and bulk water. One still wonders if there is a separate dynamic process identifiable with the hydration shell. We argue below, based on calculations of the dipolar response of hydration layers, that there is no specific shell dynamics detectable for NALMA solutions. Here we look at this problem from the perspective of the DLS susceptibility. 

One can argue that the response of the hydration layer is hard to disentangle from the background of bulk water, particularly at small concentrations represented here by C0 configuration. In order to address this question, we calculate the DLS response of only a small region of the C0 simulation cell including the solute and a water shell 5 \AA\ thick around it (see Fig.\ \ref{fig:6} for the definition of the shell). If this small region shows the dynamics distinct from the bulk, one would be able to separate out a new relaxation process. 

\begin{figure}
\includegraphics*[width=8cm]{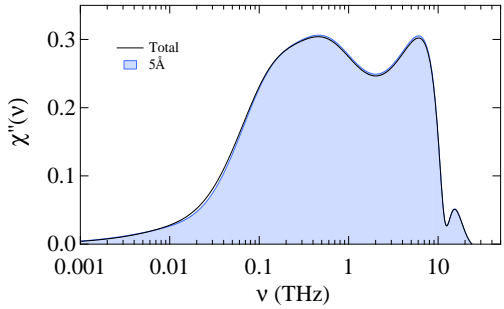}
\caption{DLS loss spectrum of the entire C0 sample (black lines) compared to the spectrum calculated for the region of the simulation box including the solute and the water shell 5 \AA\ thick (Fig.\ \ref{fig:6}) surrounding the solute (blue line and shaded area). }
\label{fig:8}
\end{figure}

Since the long-wavelength electromagnetic radiation interacts with the entire sample, the linear response of the selected region is related to the correlation function between the region polarizability $\bm{\Pi}(a,t)$ and the polarizability of the entire sample $\bm{\Pi}(0)$
\begin{equation}
\label{eq:11}
C^{\Pi}(a,t) \propto \sum_{\alpha\ne\beta}\langle \delta \Pi_{\alpha\beta}(a,t) \delta \Pi_{\alpha\beta}(0)\rangle .
\end{equation}
The result of this calculation is compared to the polarizability correlation function $C^{\Pi}(t)$ of the entire C0 sample [Eq.\ \eqref{eq:5}] in Fig.\ \ref{fig:8}. 

The fitting of the correlation functions in this case is done by using Eqs.\ \eqref{eq:8} and \eqref{eq:8-1}, with one exponent ($N_e=1$) in the cross term. Given the restrictions imposed on the correlation function, this choice implies only two fitting parameters, amplitudes $B_1$ and $B_2$ in Eq.\ \eqref{eq:8}. There is nearly no detectable difference between the response of the hydration layer and of the entire ensemble (Fig.\ \ref{fig:8}). Nevertheless, the correlation function of the C0 solution contains a negative-amplitude contribution, of about 6\% of the overall amplitude, with the Debye relaxation time of 5 ps (31 GHz). While this component might originate from cross correlations between the polarizability of NALMA and water, a signal from hydration water, which was reported to have a similar relaxation time,\cite{Lupi:2012bf} cannot be excluded. The analysis of the 5 \AA\ water layer results in a decrease of this signal by about 2\% of the entire amplitude, which, given its negative sign, implies an increase of its intensity in the spectrum. Reasons for these trends are hard to specify since we could not extract other structural and/or dynamics signatures of the hydration layer that can be attributed to this relaxation time. The conclusion that this component is a dynamic signature of solute-solvent cross-correlations\cite{Rudas2:06} seems to be most reasonable at the moment.  

\begin{figure}
\includegraphics*[width=9cm]{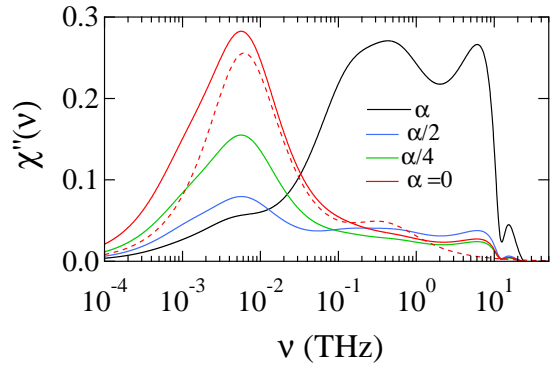}
\caption{Alteration of the DLS spectrum of C1 solution with decreasing polarizability of water. All loss functions are scaled to produce consistent normalization to their corresponding $t=0$ correlation functions (obtained by integrating $\chi''(\omega)/\omega$, see text). The dashed red line shows the DLS spectrum of a single solute in C0 solution. 
}
\label{fig:9}
\end{figure}

A relatively small amplitude of solutes' rotational relaxation in the overall DLS spectrum is the result of cancellation from cross-correlations between solute rotations and  translational motions of hydration waters induced by them.  The compensating effect of the solute-induced water motions is reduced once the polarizability of water is scaled down. This is shown in Fig.\ \ref{fig:9} where $\alpha=(1/3)\mathrm{Tr}[\bm{\alpha}]$ of water is progressively reduced. The water peak scales down and the peak of solute rotations scales up, eventually reaching the solute component of the DLS spectrum (red solid lines in Fig.\ \ref{fig:9}). All functions in Fig.\ \ref{fig:9} have been scaled to match their corresponding $t=0$ variances. Specifically, $\chi''(\omega)/\omega$ integrates [Eq.\ \eqref{eq:6}] to the $t=0$ polarizability variance $C^{\Pi}(0)$. Therefore, the polarizability loss function of the entire sample was normalized to one and $\chi''(\omega)/\omega$ of all other functions were normalized to the ratio of the corresponding variance to $C^{\Pi}(0)$.      

All calculations of the DLS spectra clearly point to a consistent picture in which the intermediate peak ($\sim 27$ ps for the second-rank polarizability response) comes from solute's rotations, without any detectable response from solute's hydration layer. A good match between the low-frequency peak in the DLS spectrum and the corresponding peak in the dielectric spectrum, assigned to solute rotations, has been recently reported.\cite{Perticaroli:2013gc} Experiment does not offer sufficient resolution to separate different components of the low frequency peak, which we can do here from MD simulations. We find that the lowest-frequency component of this composite peak is caused by solute translations producing time-dependent DID polarizability (cf.\ solid and dashed red lined in Fig.\ \ref{fig:9}).

\section{Dipolar response}
We want next to contrast depolarized light scattering to techniques probing the orientational manifold of the solution (dielectric spectroscopy, etc.). This interest is driven by the idea that the solute might exerts a more extended perturbation of the orientational structure of the hydration layer than of its density profile. The density perturbation is well recognized to be limited to the first shell only, while spontaneous polarization of the interface exists even at the water-air interface. For a sufficiently large solute carrying groups of different polarity, the hydration layer will split into domains of different polarization. Their mutual frustration can result in a considerable spatial extent of the orientational perturbation.   

\begin{figure}
\includegraphics*[width=8cm]{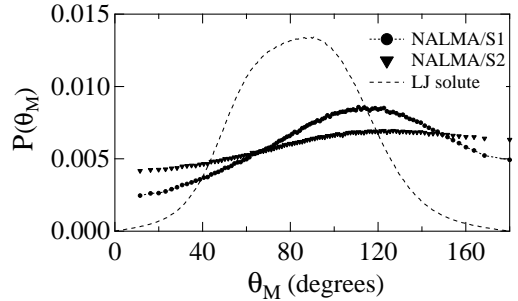}
\caption{Normalized distribution of polar angles $\theta_M$ between water dipoles and the radial direction connecting NALMA's center of mass to the oxygen of the water molecule. Reported are calculations for waters in the first (S1, $a\le4$ \AA) and the second (S2, $4\le a \le 7$ \AA)  (Fig.\ \ref{fig:6}) hydration shells of NALMA. The results for NALMA (points) are compared to a similar calculation performed for a Lennard-Jones (LJ) solute in SPC/E water (dashed line).\cite{DMjcp2:11} }
\label{fig:10}
\end{figure}

Before proceeding to a quantitative description of the orientational response of the hydration layer, one naturally wonders what kind of solute-solvent interface is produced by hydrated NALMA. It is commonly considered a hydrophobic solute, but a more quantitative metric is desirable. It is now recognized that the orientational structure of water can be quite different next to hydrophobic and hydrophilic solutes. Waters tend to orient parallel to the dividing solute-solvent surface at hydrophobic interfaces,\cite{Lee:84,McFearin:2008ys,Bratko:09} but can flip their dipoles along the local electric field when surface polar groups or charges are involved.\cite{Nihonyanagi:2009vn,Mondal:2012zr}  We therefore use here the angular distribution of interfacial water dipoles to characterize the NALMA-water interface.\cite{Pizzitutti:07} The dipole moment of waters is projected on the radial direction from NALMA's center of mass (Fig.\ \ref{fig:6}), and the distribution of these angles in the hydration layer 4 \AA\ thick is shown in Fig.\ \ref{fig:10}. 

The distribution function is broad and noticeably skewed compared to what is typically observed at the interface with a Lennard-Jones (LJ) solute\cite{DMjcp2:11} (dashed line in Fig.\ \ref{fig:10}).  The first-order orientational order parameter is $p_1=-0.13$, where for the order $\ell$ one has  $p_{\ell}=\langle P_{\ell}(\cos \theta_M)\rangle$, $P_{\ell}(x)$ is the Legendre polynomial. The preferential orientation of water dipoles is therefore toward the solute, as found for the water-air\cite{Horvath:2013fe} or water-oil\cite{Bresme:2010ch} interfaces. Correspondingly, the  second-order parameter $p_2=-0.07$ is significantly less negative than $p_2\simeq - 0.2$ typically found for waters interfacing LJ solutes\cite{DMjcp2:11} or at the water-air interface.\cite{Sokhan:97} Overall, there is  a noticeable hydrophilic character to NALMA's interface, which is not that surprising given NALMA's dipole moment of $m_0=5.7$ D.

\begin{figure}
\includegraphics*[width=8cm]{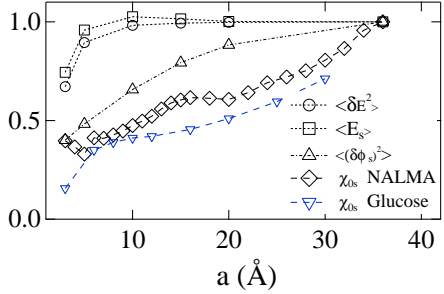}
\caption{Dependence of electrostatic observables on the thickness of the hydration layer used in the calculations. Shown are the average field projected on NALMA's dipole (Onsager's reaction field), electric field variance, and the variance of the electrostatic potential produced by the water shell. Also shown is  $\chi_{0s}(a)$ [Eq.\ \eqref{eq:20}] for NALMA calculated here and for glucose from Ref.\ \onlinecite{DMjcp3:12}. All parameters, except the glucose data, are normalized to their values at the half of the simulation box $L/2$.   
}
\label{fig:11}
\end{figure}

We now proceed to the next step, quantifying the spatial extent of the orientational structure perturbation induced by NALMA.  We have chosen to calculate several electrostatic properties typically appearing in theories of solvation and spectroscopy to look at the characteristic length-scale of the solute-induced perturbation. Figure \ref{fig:11} shows the accumulation of these parameters with increasing thickness $a$ of the shell surrounding a single NALMA molecule in C0 solution (Table \ref{tab:1}). The hydration shell in these calculations includes all waters within the distance $a$ from their oxygens to the closest point on the van der Waals surface of NALMA (Fig.\ \ref{fig:6}). 

The electrostatic properties shown in Fig.\ \ref{fig:11} are the average projection of the electric field $\mathbf{E}_s$ at the center of the NALMA molecule on its dipole moment (Onsager's reaction field\cite{Onsager:39}) and the variance of the electric field $\langle(\delta E_s)^2\rangle$. Also shown is the variance of the electrostatic potential produced by water shells at the solute $\langle (\delta \phi_s)^2 \rangle$, which gives access to the free energy of electrostatic solvation in linear response theories.\cite{Hummer:98} All these properties are calculated for waters within the shell of thickness $a$ surrounding the solute, and the resulting dependence is normalized by the corresponding values at $a=L/2\simeq 36$ \AA, where $L$ is the size of the simulation cell (Table \ref{tab:1}). Since the electrostatic field decays faster  with the distance than the electrostatic potential, it approaches its bulk value within $\sim 10$ \AA, while the electrostatic potential variance reaches saturation at $\sim 30$ \AA. Both functions demonstrate a long-range, collective behavior, with a large number of waters involved in the buildup of the electrostatic response (there are 5351 waters in the 30 \AA\ shell and 343 waters in the 10 \AA\ shell).

To get a more direct access to correlations between orientations of the solute and the surrounding waters, we have also calculated the static cross correlation function of the solute and solvent dipole moments
\begin{equation}
\chi_{0s}(a) \propto \langle \delta \mathbf{m}_0 \cdot \delta \mathbf{M}_s(a) \rangle .  
\label{eq:20}
\end{equation}
Figure \ref{fig:11} shows $\chi_{0s}(a)$, also normalized to its value at $a=L/2$. This function barely saturates on the size of the simulation box. The results obtained here are quite close to previous calculations of glucose in SPC/E water\cite{DMjcp3:12} (blue triangles in Fig.\ \ref{fig:11}). The long range of convergence of $\chi_{0s}(a)$ might have been anticipated given that $\chi_{0s}(a)$ carries a similarity with the Kirkwood factor of polar liquids, which is known to be long-ranged.\cite{SPH:81,Karlstrom:2011ht} It is therefore useful to make connection to this parameter describing orientational dipolar correlations.  

\begin{figure}
\includegraphics*[width=8cm]{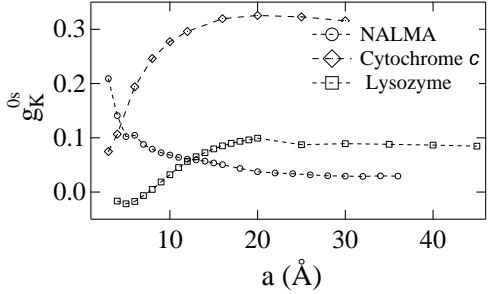}
\caption{Dependence of the solute-solvent partial Kirkwood factor [Eq.\ \eqref{eq:24}] on the thickness of the hydration layer surrounding a single NALMA solute (C0 configuration in Table \ref{tab:1}); NALMA's dipole is $m_0=5.68$ D and the dipole moment of SPC/E water is $m_s=2.35$ D. Also shown are the results for hydration shells of cytochrome \textit{c}\cite{DMjcp1:12} and lysozyme\cite{DMunpub:13} proteins at 300 K hydrated in TIP3P water.  } 
\label{fig:12}
\end{figure}

The standard definition of the Kirkwood factor of a pure homogeneous solvent is given by the relation\cite{SPH:81}
\begin{equation}
g_K = (m_s^2 N_s)^{-1} \langle (\delta M_s)^2 \rangle ,
\label{eq:22}
\end{equation}
where $m_s$ is the individual solvent (water) dipole. This definition can be extended to a mixture of two components $a$ and $b$ in terms of the partial Kirkwood factor
\begin{equation}
g_K^{ab} = \left( m_a m_b \sqrt{N_aN_b}\right)^{-1} \langle \delta \mathbf{M}_a\cdot \delta \mathbf{M}_b\rangle .
\label{eq:23}
\end{equation}
For a single solute one gets a function depending on the shell thickness
\begin{equation}
g_K^{0s}(a) = \left( m_0 m_s \sqrt{N_s(a)}\right)^{-1}\langle \delta\mathbf{m}_0\cdot \delta\mathbf{M}_s(a)\rangle . 
\label{eq:24}
\end{equation}
This function is shown in Fig.\ \ref{fig:12}. The saturation of $g_K^{0s}(a)$ to the bulk $g_K^{0s}$ is more pronounced compared to $\chi_{0s}(a)$. However, the qualitative outcome is the same.  The ``hydration layer'' defined as the region of perturbed orientational structure is much more extended into the bulk compared to the ``hydration layer'' defined as the region of perturbed local density. The density perturbation produced by NALMA is both weak and short-ranged (see Fig.\ S5 in SM\cite{supplJCP}).

The results for NALMA are compared in Fig.\ \ref{fig:12} to the same function calculated from MD simulations of two globular proteins, cytochrome \textit{c}\cite{DMjcp1:12} and lysozyme,\cite{DMunpub:13} hydrated by TIP3P waters. In contrast to enhanced correlations between the NALMA and water dipoles in the closest hydration layer, the situation is just the opposite for proteins. The reason is that waters facing charged/polar residues at the protein surface are mostly sensitive to local interactions with these residues. Correlations with the global multipole of the protein  develop only in more distant layers. Protein solvation can therefore be characterized as surface solvation, having little to do with the overall charge distribution of the solute.\cite{DMjpcl2:12} The orientational perturbation also propagates longer distance into the bulk compared to NALMA. The comparison between the hydrated NALMA dipeptide and hydrated proteins highlights the danger of extrapolating the results obtained for relatively small solutes to much larger proteins capable of both breaking the network of hydrogen bonds specific to bulk water\cite{ChandlerNature:05} and of establishing a distinctly new network of hydrogen bonds,\cite{Pizzitutti:07} with a significantly altered interfacial orientational order.

Given an extended perturbation of their orientational structure, one wonders how much the dynamics of hydration shells are perturbed. This question does not have a simple answer since solvent modes of different extent of collective behavior are reflected by different observables. Starting with a variable with a significant collective character, one can look at how the dynamics of the layer's dipole moment builds up as the shell extends in its size. 

In linear response theories, the response of a given part of the system needs to be correlated with the same variable determined for the entire region affected by the external perturbation.\cite{Hansen:03} In the case of a solution placed in a uniform electric field, the time-dependent induced dipole moment of a hydration layer can be calculated from the time-dependent correlation function\cite{DMcpl:11}
\begin{equation}
C_M(a,t) \propto \langle \delta\mathbf{M}_s(a,t) \cdot \delta \mathbf{M}_s(0) \rangle,
\label{eq:25}
\end{equation}
where $\mathbf{M}_s(t)$ refers to the total dipole moment of all waters in the sample (cf.\ to Eq.\ \eqref{eq:11}). 

\begin{figure}
\includegraphics*[width=8cm]{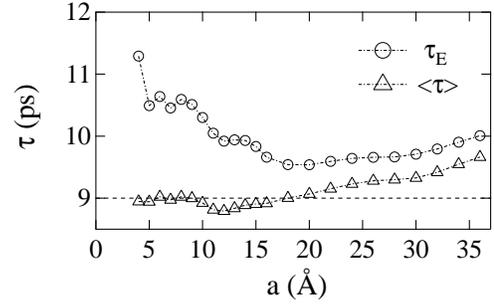}
\caption{Dynamics of the dipole moment of hydration layers of varying thickness (Fig.\ \ref{fig:6}). The relaxation times were calculated from MD simulations of C0 solution by exponential fits of the cross time correlation function $C_M(a,t)$ in Eq.\ \eqref{eq:25}. $\tau_E$ is the longest exponential relaxation time and $\langle\tau\rangle$ is the average relaxation time [Eq.\ \eqref{eq:26}]. The dashed horizontal line marks the Debye relaxation of bulk SPC/E water.\cite{Ronne:97,DMjpcb1:06}}
\label{fig:13}
\end{figure}

The results for the shell dipolar dynamics are shown in Fig.\ \ref{fig:13}. The shells increase in thickness starting from  $a=4$ \AA\ with $N_w=48$ waters in the shell. The correlation function $C_M(a,t)$ is multiexponential. Therefore, two relaxation times are shown in the plot. The time $\tau_E$ reports the slowest exponential tail of $C_M(a,t)$, while $\langle\tau \rangle $ is the average relaxation time
\begin{equation}
\langle \tau (a) \rangle = \int_0^{\infty} \left[C_M(a,t)/C_M(a,0) \right]dt .
\label{eq:26}
\end{equation}
The dashed horizontal line in Fig.\ \ref{fig:13} marks the Debye relaxation time of SPC/E water.\cite{Ronne:97,DMjpcb1:06} 

An interesting result of these calculations is a speedup of the average dipolar relaxation in the closest hydration layers due to an increasing fraction of the faster relaxation component. A similar result was reported for the water-air interface where the speedup is related to a higher number of faster rotating surface molecules with broken hydrogen bonds.\cite{Kuhne:2010uq}

It is clear from Fig.\ \ref{fig:13} that the dynamics of the shell dipole moment hardly change compared to the bulk. This is consistent with the DLS response calculations revealing no separate relaxation process attributable to the hydration shell. This conclusion is also in line with our previous simulations of SPC/E water in contact with model solutes, which showed no slowing down of the dipolar response for hydrophobic solutes of the size comparable to NALMA.\cite{DMcpl:11} Those simulations\cite{DMcpl:11} have also shown a good match between the dynamics of the shell dipole and the dynamics of the electrostatic field produced by the hydration shell inside the solute. Since the latter property is a major cause of the dynamic Stokes shift of solvated chromophores,\cite{Reynolds:96} one can anticipate that no slowing down of the Stokes shift dynamics will be recorded around hydrophobic optical dyes of the size similar to NALMA's (radius of $\sim 3.6$ \AA).

\section{Single-particle dynamics}
The results of the previous analysis indicate no dynamic signature of NALMA's hydration layer in either translational dynamics probed by the DLS susceptibility or in the collective relaxation of the shell dipole related to, but not directly accessible by, dielectric spectroscopy. One naturally wonders if this weak dynamic signature extends to single-particle rotational dynamics of water probed by the NMR.\cite{Kowalewski:2006uq} In the latter case, the relaxation function of interest is the second-order ($\ell=2$) one-particle time correlation function\cite{Laage:11,Kowalewski:2006uq,Bagno:2005kx}
\begin{equation}
C_{\ell}(t) = \left\langle P_{\ell}(\mathbf{\hat u}(t)\cdot \mathbf{\hat u}(0))\right\rangle
\label{eq:26-1}
\end{equation}
describing the dynamics of a unitary vector $\mathbf{\hat u}(t)$ of molecular orientation. NMR does not resolve the time correlation function and reports only the average time $\langle\tau_2\rangle$, i.e., the time integral of the $\ell=2$ correlation function $C_{\ell}(t)$ 
\begin{equation}
\langle \tau_{\ell} \rangle = \int_0^{\infty} C_{\ell}(t) dt .
\label{eq:27}
\end{equation}

Depending on the setup of the NMR experiment, the dynamics of different unit vectors are probed. Since the result depends on the choice,\cite{TielrooijScience:10} we present in Table \ref{tab:3} the MD calculations using either the unit dipole vector, $\mathbf{\hat e}_1$, the unit vector along the OH, $\mathbf{\hat e}_2$, and the unit vector perpendicular to the water molecular plane, $\mathbf{\hat e}_3$.  

\begin{figure}
\includegraphics*[width=8cm]{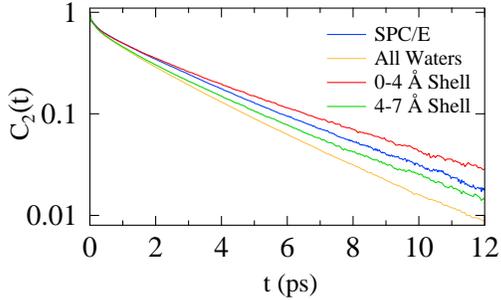}
\caption{Single-molecule rotational correlation function $C_2(t)$ [Eq.\ \eqref{eq:26-1}] calculated for SPC/E water and for C1 solution of NALMA. For the solution, averages over all waters in the simulation cell and averages over waters starting at $t=0$ in the first, $a\le 4$ \AA, and second, $4\le a\le 7$ \AA, hydration layers are shown. }
\label{fig:14}
\end{figure}

Consistent with our results for the dynamics of shell dipoles, we do not observe any major changes in the water dynamics in solution. There is in fact a slight speedup of the single-molecule orientational dynamics of the solution water compared to bulk water, which is consistent with QENS data at 1 M NALMA concentration,\cite{MalardierJugroot:2007dl} but inconsistent with the results reported by deuterium NMR of NALMA solutions at pH 4.5.\cite{Qvist:2009kx} A similar speedup of both the translational and orientational dynamics was recently reported in simulations of water between small paraffin-like plates at moderate solute-solvent dispersion attraction.\cite{Choudhury:2013dc} The speedup at intermediate solute-solvent attractions was followed by slowing down at higher attraction energies. 

The overall speedup of the orientational dynamics reported here for the entire ensemble of waters in the solution does not rule out the commonly observed slowing down of the first hydration layer\cite{Stirnemann:2011vn} 
since the slower dynamics of the first layer is counterbalanced by a speedup of outer layers containing more waters. This is shown in Fig.\ \ref{fig:14} where $C_2(t)$ decays slower for waters residing at $t=0$ in the first shell, but is faster for waters residing at $t=0$ in the second shell  (Table \ref{tab:3}).  

The overall outcome for the shell dynamics is dictated by the combination of polarity and the size of the solute. While water is less mobile near small hydrophobic groups,\cite{Post:2013zr} water next to large hydrophobic patches has a lower free energy penalty for creating hydrogen-bond defects\cite{Davis:2013if} and a higher mobility.\cite{Jamadagni:2011tc}  Differences in the single-molecule dynamics within the shells reflect orientational structural differences seen in the angle distribution functions in Fig.\ \ref{fig:10}. However, while the structural perturbation of the hydration layers can be quite significant, its dynamic signatures are fairly insignificant. 

We also list in Table \ref{tab:3} the single-particle relaxation times obtained for first-shell waters in aggregated solutions C2 and C3 (Table \ref{tab:1}). In addition to the expected slowing down, the difference in rotational relaxation times obtained for different rotation axes is worth mentioning. The rotational anisotropy becomes more pronounced for aggregated solutions, consistent with the idea of increasingly constrained water rotations.\cite{TielrooijScience:10} This rotational anisotropy can potentially be used to identify water trapped in aggregates. 

\begin{table}
\begin{ruledtabular}
\centering
\caption{First-order and second-order average relaxation times (ps) [Eq.\ \eqref{eq:27}] of single-molecule orientational dynamics of water in the bulk and in the solution. The unit vectors $\mathbf{\hat e}_i$ represent water dipole ($i=1$), OH orientation ($i=2$), and the direction perpendicular to water's molecular plane ($i=3$). }
\begin{tabular}{llccc}
$\langle \tau_{\ell}\rangle$ & System & $\mathbf{\hat e}_1$ & $\mathbf{\hat e}_2$ & $\mathbf{\hat e}_3$ \\
\hline
$\langle \tau_1\rangle$ & SPC/E & 5.4 & 5.4 & 3.4 \\
$\langle \tau_2\rangle$ & SPC/E & 1.8 & 2.1\footnotemark[1] & 1.4 \\
\hline
$\langle \tau_1\rangle$ & C1 & 4.5 & 4.3 & 2.8 \\
$\langle \tau_2\rangle$ & C1 & 1.5 & 1.7\footnotemark[2] & 1.1 \\
\hline
$\langle \tau_1\rangle$ & C1/S1\footnotemark[3] & 6.2 & 5.7 & 3.7 \\
$\langle \tau_2\rangle$ & C1/S1 & 2.1 & 2.4 & 1.7 \\
\hline
$\langle \tau_1\rangle$ & C1/S2\footnotemark[4] & 5.0 & 4.7 & 3.0 \\
$\langle \tau_2\rangle$ & C1/S2 & 1.7 & 1.9 & 1.3 \\
\hline
$\langle \tau_1\rangle$ & C2/S1 & 15.5  &  11.8  &         6.9 \\
$\langle \tau_2\rangle$ & C2/S1 & 5.9 &   5.3 &    4.9 \\
\hline
$\langle \tau_1\rangle$ & C3/S1 & 14.0 &  11.8 &   6.4 \\
$\langle \tau_2\rangle$ & C3/S1 & 5.6 &   5.2 &   4.7 \\
\end{tabular}\\
\footnotetext[1]{Experimental NMR values for reorientation times of OH and OD are 2.1 ps and 2.5 ps, respectively, see J.\ Jonas \textit{et al}. J.\ Chem. Phys. \textbf{65}, 582 (1976). 2.6 ps is reported for the rotation time of OD of HOD dissolved in HOH based on fs infrared spectroscopy measurements.\cite{TielrooijScience:10} }   
\footnotetext[2]{QENS reports 1.63 ps for the OH orientational diffusion in 1 M solution of NALMA,\cite{MalardierJugroot:2007dl} NMR of NALMA in D$_2$O at pH 4.5 and concentrations below 0.22 M reports 1.70 times slower dynamics of NALMA's first hydration layer.\cite{Qvist:2009kx} }
\footnotetext[3]{First shell of thickness $a=4$ \AA\ (Fig.\ \ref{fig:6}). }
\footnotetext[4]{Second shell defined by $4\le a \le 7$ \AA. }
\label{tab:3}
\end{ruledtabular}
\end{table}

\section{Conclusions}
In conclusion, we find that DLS response of pure water is mostly affected by its translational dynamics projected on anisotropy relaxation of the DID polarizability. The main relaxation components of NALMA solutions are bulk water and rotations of solutes with the second-order rotational diffusion time of $\simeq 27$ ps and the corresponding first-order rotational diffusion time of $\simeq 80$ ps. We have found no detectable DLS response from the hydration layer distinct from the bulk. In addition, a relatively weak signal from solute rotations is a result of cross-correlations between these rotations and water translations (density fluctuations) suppressing the rotational peak. The slowest, and relatively weak, relaxation component in the DLS response comes from collective DID polarizability caused by solutes' density fluctuations. 

This general outcome is in line with the recent emerging evidence suggesting that specific interactions are required to observe water dynamics substantially slower than bulk dynamics.\cite{Magno:2011ys,Lupi:2012bf,Vila-Verde:2013vn,Zhang:2013ly} The slowing down is therefore local.\cite{Vila-Verde:2013vn,Giammanco:2012ys,Zhang:2013ly,Nickels:2013jw} When solutes provide specific binding sites, water molecules bound to solutes demonstrate relaxation times specific to the solute motions.\cite{Wong:2012uq} Once sites for specific binding are not available, as is the case with NALMA studied here and with somewhat similar on the hydrophobicity scale t-butyl alcohol studied elsewhere,\cite{Ben-Amotz:2013vn} there is little perturbation of the water dynamics caused by the solute. For instance, recent simulations of electrolyte solutions\cite{Zhang:2013ly} have shown no slowing down of water dynamics around chloride ions, but significantly slower dynamics around stronger solvating magnesium ions, all in the same electrolyte solution.        

The conclusions reached in this study disagree with EDLS experiments, which indicate a slow, 30--50 GHz, component of water dynamics both for sugar\cite{Lupi:2012bf} and NALMA\cite{Petricaroli:11} solutions. Plausible sources of discrepancy between simulations and experiments may be envisaged in the force-field and in the perturbation approximation adopted for the solution polarizability. Deprotonation equilibria, with slow water located around ionized sites, cannot be excluded as well.  
  
While no specific response of the hydration layer is seen from the density fluctuations of hydration waters, there is a substantial orientational structure of the hydration layer, which propagates $\simeq 3-5$ water layers into the bulk. It is reflected in a slow saturation of electrostatic observables with increasing size of the hydration shell and in a similarly slow decay of dipole-dipole correlations (partial Kirkwood factor). 

The orientational perturbation of the hydration layers does not result in a significant dynamic signature. The changes of the dynamics of the shell dipole moment and of single-molecule rotations are fairly insignificant. This latter outcome, and the lack of a significant structural perturbation of the second hydration shell, are not necessarily universal results and might depend on both the size of the solute and the distribution of surface polarity.\cite{DMjpcl2:12} Proteins, which are both larger in size and possess a higher density of the surface charge, clearly show slowing down of the hydration shell dynamics, both in the density and orientational manifolds.\cite{Jordanides:99,Zhang:07,Bhattacharyya:2008fk,DMjcp1:12,Bagchi:2012hc} 

There is a clear distinction between the perturbation produced by the solute to the interfacial structure and to the interfacial dynamics. Potential energies affect structure, while forces determine the dynamics. Interaction potentials, in particular electrostatic interactions, can be long-ranged and significant (compared to $k_\text{B}T$), while forces produced by these interactions can be weak relative to local forces affecting local dynamics. Short-ranged specific interactions, producing stronger forces, play a more essential role in affecting the dynamics, while long-ranged interactions are responsible for the structure of the interface. 

From the technical side, our calculations stress on the necessity to correctly calculate the collective dynamic and statistical properties of a chosen sub-ensemble, hydration shells in this study. Linear response of a sub-ensemble to an external perturbation necessarily includes cross-correlations with the same property determined for the entire region affected by the perturbation (the entire simulation box for long-wavelength radiation). Cross-correlations are significant both dynamically and statically and cannot be neglected.\cite{Rudas2:06,DMjcp1:12}

\acknowledgments This research was supported by the National Science
Foundation (CHE-1213288). CPU time was provided by the National Science Foundation through XSEDE resources (TG-MCB080116N). 

%\bibliography{dpls,chem_abbr,dielectric,dm,statmech,protein,liquids,solvation,dynamics,elastic,simulations,surface,nano,water,glass,et}


\end{document}